\documentclass[cameraready]{Interspeech}

\definecolor{arytenoid-cartilage}{rgb}{0.54, 0.17, 0.89} 
\definecolor{epiglottis}{rgb}{0.25, 0.88, 0.82} 
\definecolor{lower-lip}{rgb}{0.00, 1.00, 0.00} 
\definecolor{pharyngeal-wall}{rgb}{0.85, 0.66, 0.13} 
\definecolor{soft-palate-midline}{rgb}{0.12, 0.56, 1.00} 
\definecolor{tongue}{rgb}{1.00, 0.55, 0.00} 
\definecolor{upper-lip}{rgb}{1.00, 0.00, 1.00} 
\definecolor{vocal-folds}{rgb}{1.00, 0.41, 0.71} 
\usepackage{float}
\usepackage{placeins}
\usepackage{stfloats}

\usepackage{todonotes}

\newif\ifshowtodos

\title{Acoustic-to-Articulatory Inversion of Clean Speech Using an MRI-Trained Model}

\author[affiliation={1}, orcid=0009-0002-2907-7477]{Sofiane}{Azzouz}
\author[affiliation={2}, orcid=0000-0003-3519-1743]{Pierre-André}{Vuissoz}
\author[affiliation={1}, orcid=0000-0002-2379-6481]{Yves}{Laprie}

\address{
    $^1$ {Université de Lorraine, CNRS, Inria, F-54000 Nancy, France}\\
    $^2$ {Université de Lorraine, Inserm, IADI U1254, F-54000 Nancy, France}
}

\email{sofiane.azzouz@loria.fr, pa.vuissoz@chru-nancy.fr, yves.laprie@loria.fr}

\usepackage{comment}

\begin{document}

\maketitle
\begin{abstract}
Articulatory acoustic inversion reconstructs vocal tract shapes from speech. Real-time magnetic resonance imaging (rt-MRI) allows simultaneous acquisition of both the acoustic speech signal and articulatory information. Besides the complexity of rt-MRI acquisition, the recorded audio is heavily corrupted by scanner noise and requires denoising to be usable. For practical use, it must be possible to invert speech recorded without MRI noise.
In this study, we investigate the use of speech recorded in a clean acoustic environment as an alternative to denoised MRI speech. To this end we compare two signals from the same speaker with identical sentences which are aligned using phonetic segmentation.
A model trained on denoised MRI speech is evaluated on both denoised MRI and clean speech. We also assess a model trained and tested only on clean speech. Results show that clean speech supports articulatory inversion effectively, achieving an RMSE of 1.56 mm, close to MRI-based performance.
\end{abstract}


\section{Introduction}
Acoustic-to-articulatory inversion aims to estimate the geometric shape of the vocal tract from the acoustic speech signal. 
Since the 1970s, several families of solutions have been proposed to address this complex inverse problem. 

With the availability of measured articulatory data, such as Electromagnetic Articulography (EMA) \cite{Wrench2000} and X-ray Microbeam (XRMB) \cite{westbury1990x}, it became possible to establish a direct relationship between the acoustic signal and vocal tract geometry. Statistical approaches emerged, including Gaussian Mixture Models (GMM) \cite{toda2004acoustic} and Hidden Markov Models (HMM) \cite{hiroya2004estimation}. Later, neural networks significantly improved performance, with architectures such as Bi-LSTM \cite{liu2015deep,parrot2020}, Bi-GRNN \cite{wu2023speaker}, and Temporal Convolutional Networks (TCN) \cite{siriwardena2023secret}. 


Real-time Magnetic Resonance Imaging (rt-MRI) has been introduced as an alternative to EMA, as it overcomes many of its limitations and constraints by enabling visualization of the entire vocal tract from the glottis to the lips. A first MRI corpus was described in \cite{ramanarayanan2018analysis}; however, it is limited by relatively low image resolution (68 × 68 pixels) and by heavily corrupted audio due to scanner noise. 

Some studies have directly applied inversion to MRI images \cite{Csapo2020, Oura2024}, while others have proposed generative models, including diffusion-based approaches \cite{nguyen2025speech2rtmri}. In both cases, performance is typically evaluated at the pixel level of the reconstructed images. An alternative approach consists of automatically segmenting MRI images to extract articulator contours, as proposed in \cite{ribeiro:hal-04376938}, which achieves very promising results. In this framework, inversion is performed on articulatory contours rather than raw MRI images. Following this direction, \cite{azzouz:hal-05292817} successfully inverted the tongue only, and \cite{azzouz:hal-05293831} extended this approach to reconstruct the complete vocal tract. Furthermore, the use of higher-resolution images (136 × 136 pixels), combined with better-denoised audio, significantly improves data quality. These aspects will be detailed in Section \ref{dataset}.

The majority of previous works use Mel-Frequency Cepstral Coefficients (MFCCs) or Mel-Generalized Cepstral Line Spectral Pairs (MGC-LSP) \cite{Csapo2020} as input features. More recently, Self-Supervised Learning (SSL) methods have demonstrated remarkable performance in acoustic-to-articulatory inversion. In particular, HuBERT \cite{hsu2021hubert} stands out \cite{cho2024self}, outperforming approaches such as wav2vec 2.0 \cite{baevski2020wav2vec}, WavLM \cite{chen2022wavlm}, as well as classical MFCCs features. Recent studies further confirm the superiority of HuBERT on XRMB data \cite{attia2024improving}. Moreover, these representations have proven effective in tasks such as speaker verification \cite{wang2021fine}, which is closely related to the objectives of the present study.

\begin{figure*}[!b] 
  \centering
  \includegraphics[width=1\textwidth]{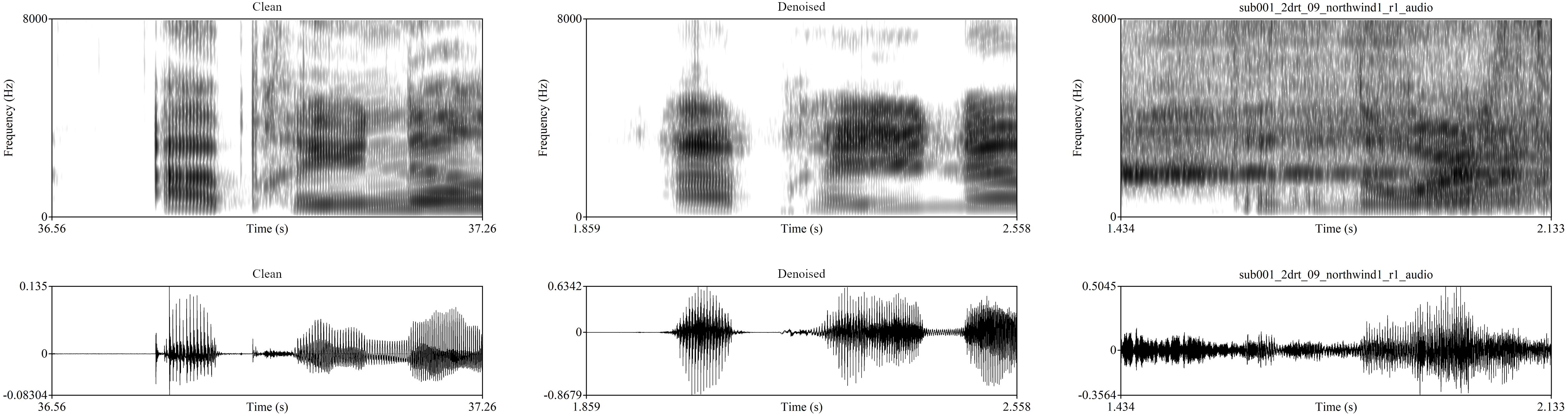}
  \caption{
    Comparison of normal speech, MRI-denoised speech from our dataset, and denoised speech from \cite{ramanarayanan2018analysis}. The first two signals were produced by the same female speaker uttering “Après une heure.” All three audio files are provided as supplementary material.
  }
  \label{fig:spectrogram_comparison} 
\end{figure*}

In this work, we focus on the impact of acoustic signal quality on articulatory inversion, particularly on the transition from noisy speech (recorded in the MRI environment) to clean speech. We use an rt-MRI database that shares similar characteristics with the one described in \cite{isaieva:hal-03507532}, but with higher image resolution (136 × 136 pixels). Our objective is to evaluate the effect of using clean speech instead of noisy speech on the performance of a model trained on MRI data. To ensure temporal alignment between the two types of signals, phonetic alignment is applied. We analyze the impact of this substitution either on the test data only or on both the training and test sets.

\section{Impact of MRI noise and denoising}
Our goal is to be able to use articulatory acoustic inversion in real-world applications, which means that the inversion must be able to use speech recorded in a quiet environment. However, speech recorded in an MRI machine, even if it is denoised before use, is still far from speech produced in a quiet environment. Figure~\ref{fig:spectrogram_comparison} illustrates the difference between a signal recorded in a quiet room and two speech signals recorded in an MRI machine and then denoised, the first from the database we use in this work and the second from the database collected by \cite{ramanarayanan2018analysis}.
It is clear that the denoising we used gives much better results, but despite this, the denoised signal is still quite different from normal speech (the energy is lower in the high frequencies in particular). It is therefore necessary to adapt the inversion system so that the results remain relevant.

The second difference between normal speech and speech produced in an MRI machine comes from the Lombard effect \cite{junqua1996influence} and the supine position. We do not take this effect into account because we do not have the data to address this issue, as it is currently impossible to have an MRI machine in a vertical position.

\section{Dataset} \label{dataset}
We used two different corpus recorded by the same native French female speaker. The first corpus was recorded at the Centre Hospitalier Régional Universitaire (CHRU) de Nancy and contains approximately 2.5 hours of data. It consists of 105 acquisitions, each lasting 80 seconds and containing 4000 MRI frames. The rt-MRI images have a spatial resolution of 136 × 136 pixels, with a temporal resolution of 50 ms per frame (20 fps). The in-plane pixel size is 1.62 mm. The corresponding audio was recorded using an optical microphone
at a sampling rate of 16 kHz and was subsequently denoised using \cite{Ozerov2012}. In addition, we have a phonetic segmentation into 44 phonemes. The alignment was first performed automatically using Astali \cite{fohr2015importance} and then manually corrected by an expert.
The second corpus was recorded in a clean environment, outside the MRI scanner. The same speaker was recorded while producing the same set of sentences as in the first dataset. The audio was initially recorded at a sampling rate of 48 kHz and then downsampled to 16 kHz to match the MRI dataset. In addition, phonetic segmentation was automatically generated using the Montreal Forced Aligner \cite{mcauliffe2017montreal} and subsequently manually corrected by an expert. Both phonetic segmentations were carefully reviewed to ensure they were strictly identical across the two corpora.



\subsection{Pre processing}

We compared the HuBERT Base and Large models, the best results were obtained with the Base model. Therefore, in this study, we use only HuBERT-Base representations as input embeddings. HuBERT processes 16 kHz audio and outputs 768-dimensional embeddings at a frame rate of 50 Hz.

\begin{figure}[H] 
  \centering
  \includegraphics[width=0.471\textwidth]{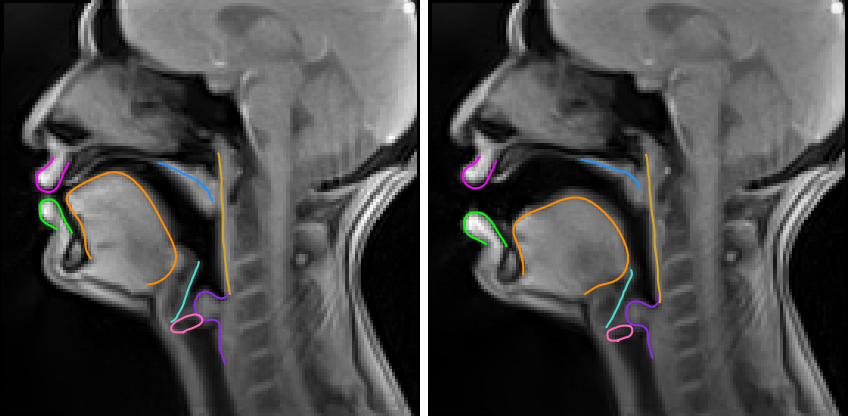} 
  \caption{
    Segmentation of articulators contour tracked in two images of the rt-MRI film:\hspace{0.1cm}
    \raisebox{0.25ex}{\textcolor{arytenoid-cartilage}{\rule{0.18cm}{0.08cm}}} Arytenoid cartilage,
    \raisebox{0.25ex}{\textcolor{epiglottis}{\rule{0.18cm}{0.08cm}}} Epiglottis,\\
    \raisebox{0.25ex}{\textcolor{lower-lip}{\rule{0.18cm}{0.08cm}}} Lower lip,
    \raisebox{0.25ex}{\textcolor{pharyngeal-wall}{\rule{0.18cm}{0.08cm}}} Pharyngeal wall,
    \raisebox{0.25ex}{\textcolor{soft-palate-midline}{\rule{0.18cm}{0.08cm}}} Soft palate midline,
    \raisebox{0.25ex}{\textcolor{tongue}{\rule{0.18cm}{0.08cm}}} Tongue,
    \raisebox{0.25ex}{\textcolor{upper-lip}{\rule{0.18cm}{0.08cm}}} Upper lip,
    \raisebox{0.25ex}{\textcolor{vocal-folds}{\rule{0.18cm}{0.08cm}}} Vocal folds
  }
  \label{fig:original_contours} 
\end{figure}

To segment the articulatory contours, we employed an automatic contour-tracking approach \cite{ribeiro:hal-04376938} and divided the vocal tract into eight articulators: the arytenoid cartilage, epiglottis, lower lip, pharyngeal wall, soft palate midline (velum), tongue, upper lip, and vocal folds (glottis) (see Figure~\ref{fig:original_contours}). Each articulator contour consists of 50 points.

\section{Methods}
\subsection{Alignment of MRI and Clean Speech Corpora}
\label{Alignement}

It was necessary to align the MRI corpus with the clean speech corpus. After manually verifying the phonetic segmentations and ensuring their consistency, we developed a hierarchical alignment algorithm based on phonetic information.

The alignment was performed hierarchically, proceeding from sentences to words, and then to phones. For each sentence in the MRI corpus, we identified the most similar sentence in the clean corpus using a similarity measure based on the Gestalt pattern matching algorithm~\cite{ratcliff1988pattern}. This algorithm finds the longest matching subsequences between two character strings to provide an
overall similarity score. Only sentences with a similarity score of 75\% or higher were considered for alignment, to account for possible pronunciation errors or repetitions by the speaker. Since each sentence is unique in the MRI corpus, at most one candidate reaches this threshold, ensuring a one-to-one correspondence.

Once corresponding sentences were identified, alignment was performed at the word and phone levels. For each word in the MRI corpus, all candidate matches in the clean corpus were selected based on textual equality. In the case of duplicates, for example when the same word appears multiple times in a sentence, we used a relative position calculation to select the most coherent match. The relative position of a word within its sentence is defined as:
\begin{equation}
    r^{\text{MRI}}_{\text{word}} = \frac{t^{\text{MRI}}_{w,\text{start}} - t^{\text{MRI}}_{s,\text{start}}}{t^{\text{MRI}}_{s,\text{end}} - t^{\text{MRI}}_{s,\text{start}}}
\end{equation}
where $t^{\text{MRI}}_{w,\text{start}}$ is the start time of the word, and $t^{\text{MRI}}_{s,\text{start}}$,
$t^{\text{MRI}}_{s,\text{end}}$ are the start and end times of the corresponding sentence in the MRI corpus. The corresponding clean word is chosen as the one whose relative position within its sentence is closest to $r^{\text{MRI}}_{\text{word}}$.

To account for differences in duration between MRI and clean phones, we applied local temporal normalization at the phone level. For each MRI frame whose midpoint falls at time $t^{\text{MRI}}_{f,\text{mid}}$ within a given phone, the duration of that MRI phone is calculated as:
\begin{equation}
    d^{\text{MRI}}_{p} = \max\left(t^{\text{MRI}}_{p,\text{end}} - t^{\text{MRI}}_{p,\text{start}},\ \epsilon\right)
\end{equation}
with $\epsilon = 0.001$\,s to prevent division by zero, where $t^{\text{MRI}}_{p,\text{start}}$ and $t^{\text{MRI}}_{p,\text{end}}$ are the start and end times of the phone in the MRI corpus. The intra-phone relative position of the MRI frame is then:
\begin{equation}
    r^{\text{MRI}}_{\text{intra}} = \frac{t^{\text{MRI}}_{f,\text{mid}} - t^{\text{MRI}}_{p,\text{start}}}{d^{\text{MRI}}_{p}}
\end{equation}
This proportion is then used to determine the target time in the corresponding phone of the clean corpus:
\begin{equation}
    t^{\text{clean}}_{p,\text{target}} = t^{\text{clean}}_{p,\text{start}}
    + r^{\text{MRI}}_{\text{intra}} \cdot \left(t^{\text{clean}}_{p,\text{end}} - t^{\text{clean}}_{p,\text{start}}\right)
\end{equation}
where $t^{\text{clean}}_{p,\text{start}}$ and $t^{\text{clean}}_{p,\text{end}}$ are the start and end times of the corresponding phone in the clean corpus. Finally, this target time is converted to a frame index for the clean signal:
\begin{equation}
    \text{idx}^{\text{clean}}_{p} = \left\lfloor \frac{t^{\text{clean}}_{p,\text{target}} \cdot f_s}{\Delta t_{\text{frame}}} \right\rfloor
\end{equation}
where $f_s$ is the sampling frequency and $\Delta t_{\text{frame}}$ is the duration of a single frame. This phone-by-phone time-stretching procedure ensures that each MRI frame is aligned with the corresponding time point in the clean signal, while preserving the phonetic structure and compensating for differences in phone duration between the two corpora.




\begin{table*}[b]
\caption{Comparison of RMSE (mm) and Median (mm) between the M2M, M2C, and C2C conditions}
\label{tab:model_comparison}
\centering

\setlength{\tabcolsep}{10pt}    

\begin{tabular}{lcccccc}
\hline
\multicolumn{1}{c}{} & \multicolumn{2}{c}{\textbf{M2M}} & \multicolumn{2}{c}{\textbf{M2C}} & \multicolumn{2}{c}{\textbf{C2C}} \\
 \multicolumn{1}{c}{} & \textbf{RMSE} & \textbf{Median} & \textbf{RMSE} & \textbf{Median} & \textbf{RMSE} & \textbf{Median} \\


\hline
\textbf{Arytenoid cartilage} & 1.72 $\pm$ 1.06 & 1.46 & 1.83$^{*}$ $\pm$ 1.12 & 1.55 & 1.78$^{**}$ $\pm$ 1.06 & 1.53 \\
\textbf{Epiglottis}          & 1.51 $\pm$ 0.83 & 1.33 & 1.71$^{*}$ $\pm$ 0.98 & 1.50 & 1.63$^{**}$ $\pm$ 0.94 & 1.42 \\
\textbf{Lower lip}           & 1.44 $\pm$ 0.77 & 1.29 & 1.57$^{*}$ $\pm$ 0.88 & 1.37 & 1.48$^{**}$ $\pm$ 0.82 & 1.30 \\
\textbf{Pharyngeal wall}     & 1.07 $\pm$ 0.54 & 0.98 & 1.17$^{*}$ $\pm$ 0.65 & 1.03 & 1.12$^{**}$ $\pm$ 0.62 & 0.99 \\
\textbf{velum}               & 1.39 $\pm$ 0.66 & 1.27 & 1.39$\phantom{^{*}}$ $\pm$ 0.68 & 1.26 & 1.30$^{**}$ $\pm$ 0.62 & 1.18 \\
\textbf{Tongue}              & 2.29 $\pm$ 1.12 & 2.07 & 2.61$^{*}$ $\pm$ 1.34 & 2.31 & 2.46$^{**}$ $\pm$ 1.22 & 2.19 \\
\textbf{Upper lip}           & 1.14 $\pm$ 0.45 & 1.08 & 1.14$\phantom{^{*}}$ $\pm$ 0.47 & 1.07 & 1.09$^{**}$ $\pm$ 0.45 & 1.02 \\
\textbf{Vocal folds}         & 1.51 $\pm$ 0.72 & 1.39 & 1.66$^{*}$ $\pm$ 0.95 & 1.45 & 1.61$^{**}$ $\pm$ 0.92 & 1.42 \\
\hline
\textbf{Mean}                & 1.51 $\pm$ 0.87 & 1.33 & 1.64$^{*}$ $\pm$ 1.02 & 1.39 & 1.56$^{**}$ $\pm$ 0.96 & 1.33 \\
\hline
\end{tabular}

\caption*{\centering
\small
$^{*}$Significant difference compared to the M2M result ($p < 0.05$) based on a t-test.\par
$^{**}$Significant difference compared to the M2M and M2C result ($p < 0.05$) based on a t-test.
}
\end{table*}


\subsection{Model Architecture}
\begin{figure}[ht]
    \centering
    \includegraphics[width=0.47\textwidth]{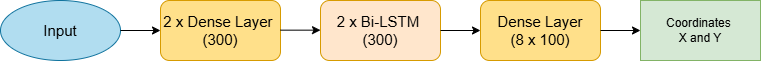} 
    \caption{Model architecture}
    \label{fig:model_architecture}
\end{figure}

The model used in this study is inspired by \cite{azzouz:hal-05292817, azzouz:hal-05293831} (see Figure \ref{fig:model_architecture}), with audio embeddings as input. It consists of five layers: two fully connected layers with 300 units each, two Bi-LSTM layers with 300 units each, and a final dense layer producing a tensor of size $8 \times 100$. This output corresponds to eight articulators, where 100 represents the contour coordinates (50 X and 50 Y coordinates), resulting in 50 points per articulator.

\subsection{Loss function}
Articulatory acoustic inversion is a regression task. In this study, the objective is to minimize the distance between the predicted and reference articulatory points. We used the Mean Squared Error (MSE) loss function, which is the most commonly used metric for such regression problems.
\begin{equation}
\text{MSE} = \frac{1}{n} \sum_{i=1}^{n} \left( y_i - \hat{y}_i \right)^2\label{eq2}
\end{equation}
where n represents the total number of observations in the dataset, \(y_i\) and \(\hat{y}_i\) represent the true and predicted values of the output for example \(i\), respectively.

\subsection{Evaluation of the model}
We denormalized the contours and removed all silence segments, including those occurring between words. We evaluated the model using the RMSE and the median error between the original and predicted denormalized contours. First, both error metrics were computed for each frame and each contour. Then, the global mean was calculated across all frames and contours. Both errors are expressed in millimeters.

\begin{equation}
\text{RMSE} = \sqrt{\frac{1}{n} \sum_{i=1}^{n} \left( y_i - \hat{y}_i \right)^2}\label{eq4}
\end{equation}

\subsection{Experiments}
The first experiment aims to compare three experimental configurations in order to assess the impact of using clean speech. In all settings, the target output remains the same: the MRI contours. The three settings are as follows:




 \begin{enumerate}


\item Input for train and eval is denoised MRI speech (M2M).
\item Input for train is denoised MRI speech, and for eval is clean speech (M2C).
\item Input for train and eval is clean speech (C2C).

\end{enumerate}


For all configurations involving clean speech, the signals were first aligned using the phonetic alignment procedure described in the section~\ref{Alignement}.


The second experiment aims to evaluate the impact of the proposed alignment method by comparing it with an approach based on Dynamic Time Warping (DTW) \cite{muller2015fundamentals}. In this setting, clean speech was aligned using DTW, and the two configurations involving clean speech, namely M2C and C2C, were reproduced using DTW-based alignment. These configurations are referred to as M2C-DTW and C2C-DTW.

\subsection{Model parameters}
Our dataset contained 412,000 images. After removing silence segments between sentences, we randomly split the data by acquisition into 80\% for training, 10\% for validation, and 10\% for testing. All models were then trained for up to 300 epochs with a batch size of 10 using the Adam optimizer with a learning rate of 0.001. Early stopping with a patience of 10 epochs was applied based on validation performance, and training was stopped when no further improvement was observed.

\section{Results}

\begin{table*}[th]
\caption{Results of RMSE (mm) and Median (mm) of models trained with alignment DTW}
\label{tab:alignment_comparison}
\centering
\setlength{\tabcolsep}{10pt}    
\begin{tabular}{lcccc}
\hline
\multicolumn{1}{c}{}  & \multicolumn{2}{c}{\textbf{M2C-DTW}} & \multicolumn{2}{c}{\textbf{C2C-DTW}} \\
 \multicolumn{1}{c}{} & \textbf{RMSE} & \textbf{Median} & \textbf{RMSE} & \textbf{Median} \\


\hline
\textbf{Arytenoid cartilage} & 1.92 $^{*}$ $\pm$ 1.21 & 1.61 & 1.90$^{**}$ $\pm$ 1.17 & 1.61 \\
\textbf{Epiglottis}          & 1.78 $^{*}$ $\pm$ 1.10 & 1.53 & 1.73$^{**}$ $\pm$ 1.01 & 1.51 \\
\textbf{Lower lip}           & 1.64 $^{*}$ $\pm$ 0.99 & 1.42 & 1.60$^{**}$ $\pm$ 0.97 & 1.37 \\
\textbf{Pharyngeal wall}     & 1.21 $^{*}$ $\pm$ 0.75 & 1.06 & 1.20$^{**}$ $\pm$ 0.66 & 1.06 \\
\textbf{velum}               & 1.50 $^{*}$ $\pm$ 0.78 & 1.34 & 1.46$^{**}$ $\pm$ 0.76 & 1.31 \\
\textbf{Tongue}              & 2.65 $^{*}$ $\pm$ 1.60 & 2.28 & 2.66$^{**}$ $\pm$ 1.52 & 2.33 \\
\textbf{Upper lip}           & 1.21 $^{*}$ $\pm$ 0.50 & 1.12 & 1.17$^{**}$ $\pm$ 0.48 & 1.09 \\
\textbf{Vocal folds}         & 1.77 $^{*}$ $\pm$ 1.08 & 1.54 & 1.72$^{**}$ $\pm$ 0.95 & 1.53 \\ \hline
\textbf{Mean}                & 1.71 $^{*}$ $\pm$ 1.13 & 1.45 & 1.68$^{**}$ $\pm$ 1.08 & 1.43 \\
\hline
\end{tabular}

\caption*{\centering
\small
$^{*}$Significant difference compared to the M2C result ($p < 0.05$) based on a t-test.\par
$^{**}$Significant difference compared to the C2C result ($p < 0.05$) based on a t-test.
}
\end{table*}

Table~\ref{tab:model_comparison} compares the results obtained with the three configurations: M2M, M2C, and C2C. The M2M configuration remains the best, with an average RMSE of 1.51 mm and a median of 1.33 mm. It is followed by C2C, which achieved an average RMSE of 1.56 mm and a median of 1.33 mm. Finally, the M2C model obtained an average RMSE of 1.64 mm and a median of 1.39 mm. Similar performance trends are observed when considering each articulator individually.

Table~\ref{tab:alignment_comparison} presents the results obtained using DTW as the alignment method between denoised MRI speech and clean speech. The results show that the M2C-DTW configuration achieved an average RMSE of 1.71 mm and a median of 1.45 mm. For individual articulators, RMSE values ranged from 1.21 mm to 2.65 mm and median values from 1.06 mm to 2.28 mm. The C2C-DTW model achieved an overall average RMSE and median of 1.68 mm and 1.43 mm, respectively, while individual articulator RMSE values ranged between 1.20 mm and 2.26 mm and median values from 1.06 mm to 2.33 mm.


\section{Discussion}
The results shown in Table~\ref{tab:model_comparison} indicate that M2M configuration achieves the best scores, both globally and per articulator. This configuration corresponds to that of our previous work on inversion, using denoised speech for both training and testing. There is therefore no mismatch between the training and test data, which explains why this configuration gives the best results.
However, if we test the system trained with denoised speech on clean speech (M2C configuration) without any adaptation, which could correspond to the use of inversion in a real-world application, the results deteriorate significantly, with the average error increasing from 1.51 mm to 1.64 mm. Nevertheless, these results are acceptable, probably because the denoised speech is of fairly good quality and not too far removed from natural speech (see Figure~\ref{fig:spectrogram_comparison}).
The system trained and tested on clean speech (C2C configuration) gives significantly better results than the M2C configuration, going from 1.64 mm to 1.56 mm. This shows the relevance of the alignment between MRI data and the clean speech signal presented in section~\ref{Alignement}. The key point is that the performance is very close to that of the original system (M2M) and that the inversion system can therefore be used on clean speech, which corresponds to a real-world application. (A video showing inverted contours generated with the C2C configuration, provided as supplementary material.)

The results presented in Table~\ref{tab:alignment_comparison}  provide a better understanding of the impact of the alignment procedure described in Section 4.1. Another solution could have been to align the acoustic signals using DTW without taking phonetic segmentation into account. We can see that both configurations (M2C-DTW and C2C-DTW) produce significantly worse results than the corresponding configurations (M2M and C2C). This confirms that using phonetic segmentation to align articulatory data with acoustic data as closely as possible offers a real advantage. Indeed, phonetic alignment takes phoneme boundaries into account: each frame is assigned to its corresponding phoneme. The alignment is therefore guided by the phonetic content rather than by acoustic similarity alone, which improves its accuracy.

\section{Conclusion}
In this paper, we studied the possibility of reconstructing the shape of the vocal tract from clean speech rather than speech recorded in an MRI scanner and then denoised. Although the results obtained are slightly inferior to the reference system trained and tested on denoised speech, they are nevertheless very satisfactory with an average error of 1.56 mm, given the spatial resolution of MRI images, which is 1.62 mm per pixel. This shows that it is now possible to consider the use of articulatory acoustic inversion in real-world applications.



\clearpage

\section{Generative AI Use Disclosure}
Artificial intelligence tools were used solely for grammatical correction and language polishing of the manuscript. All aspects concerning the scientific content, experiments, analysis, and interpretation were carried out by the authors. All authors take full responsibility for the content of this paper.

\bibliographystyle{IEEEtran}
\bibliography{mybib}
\end{document}